\documentstyle[pre,twocolumn,aps,psfig]{revtex}

\begin{document}

\title{Detecting Determinism in High Dimensional Chaotic Systems}

\author{G. J. Ortega}
\address{Centro de Estudios e Investigaciones,
Universidad Nacional de Quilmes,\\ R. S. Pe\~na 180, 1876, Bernal, Argentine}
\author{C. Degli Esposti Boschi and E. Louis}
\address{Departamento de F\'{\i}sica Aplicada and Unidad Asociada
of the Consejo Superior de Investigaciones Cient\'{\i}ficas, \\
Facultad de Ciencias, Universidad de Alicante, Apartado 99, E-03080, 
Alicante, Spain}

\date{\today}

\maketitle

\begin{abstract}
A method based upon the statistical evaluation of the differentiability
of the measure along the trajectory is used to identify determinism
in high dimensional systems. The results  show that the method is
suitable for discriminating stochastic from deterministic systems even
if the dimension of the latter is as high as 13. 
The method is shown to succeed in identifying determinism in  
electro--encephalogram signals simulated by means of a high dimensional
system. 
\end{abstract}

\indent

\section{Introduction}
\label{S_Intro}

Although numerous methods have been developed in recent years addressed
to detect determinism in time series \cite{SM90,KG92,WB93,SC94,BK99}, 
most of them give wrong or ambiguous results when  applied to high
dimensional systems (we refer to systems with attractor's dimension 
greater than, say, 
five).  However, high dimensional systems are
ubiquitous in nature, as for example in the case of spatially extended
systems. Therefore it remains the task to develop a
robust method capable of detecting deterministic behavior in systems
with many active degrees of freedom \cite{KS97}.  

We hereby claim that, contrary to some widely accepted premises
in the chaos community \cite{AB93,CF00,JK99}, it is possible to
discriminate determinism from stochasticity in systems with
a high correlation (or information) dimension. 
Our method is based upon a fundamental
property of  deterministic systems, namely, the differentiability
of the measure along the trajectory. Our numerical implementation
of this property is robust enough to uncover this topological property 
in short data sets.

As a very first step, we should clarify what
we mean by determinism. Attaching to definitions, saying that
a time series stems from a deterministic system only if it can
be produced by a discrete- or continuous-time dynamics without
any stochastic source, would isolate a very small class
of phenomena. In fact, even artificial time series from
non-stochastic dynamical systems are
typically affected by numerical noise, which at best
can be considered to be in its zero-limit.
Unfortunately, numerical errors tend to be magnified
with the reconstruction process and can mimic the
effect of a stochastic feedback.
It is 
pertinent to note that Casdagli {\it et al.} \cite{CE91} have shown
that univariate time series derived from sufficiently high dimensional 
systems cannot in practice be regarded as noise-free.
 
On the other hand, the importance of detecting a possible deterministic
dynamics is not merely academic and it is often motivated by
the intention of controlling the non-stochastic part of
the system under investigation. Thus, operationally, we prefer to
imagine an underlying dynamics composed by a stochastic part 
(that is, dynamical noise) and
a non-stochastic one. In this sense, it's not always obvious how
to measure the weight of one part with respect to the other.
However, it seems very natural to classify as {\it purely
stochastic} a system which has not residual (non-trivial)
dynamics if its stochastic part is switched off.
For example, this is the case of Eqs. (\ref{stoc1}) and (\ref{stoc2})
that will be discussed later on. More generally, we speak of
determinism whenever the dynamics has a relevant non-stochastic
backbone, being conscious that we can decide if this is really
non-negligible only when we adopt a suitable indicator.
In this paper we make use of the statistical differentiability
of the measure as introduced in refs. \cite{OL98,OL00}. There
it is argued that this quantity is sensible to the amount
of dynamical noise intrinsic to the system. Here we explore
the two opposite sides of the scenario depicted above,
showing that it sharply discriminates purely stochastic
systems from deterministic dynamics, even when the latter
takes place on rather high dimensional attractors.   

\section{Methods}
\label{S_Met}

\subsection{Tracking the measure along the trajectory}
\label{Ss_Tmat}

To illustrate the point, let's
consider a dissipative dynamical system described by $n$-first-order
differential equations ${\bf {\dot x}} = {\bf F}({\bf x})$.
The corresponding flow ${\it f}^t$ maps a "typical" initial
condition ${\bf x}_0$ into ${\bf x}(t) = {\it f}^t ({\bf x}_{0})$
at time $t$. Once transients are over, the motion settles
over the attractor ${\cal A}$. Defined over this set is the
natural measure, which, from an operational point of view, can
be considered as the limiting distribution of almost all
starting initial conditions, that is,
\begin{equation}
\mu(B_r ({\bf x}))  = \lim_{t \rightarrow \infty}
{1 \over t} \int_0^t 1_{B_r} (f^{\tau}({\bf y}_0)) \: {\rm d}\tau
\label{expr_mu}
\end{equation}
for almost all ${\bf y}_0$ in the basin of attraction.
In Eq. (\ref{expr_mu}) $B_r(\bf{x})$ indicates the hyper-sphere of radius
$r$ centered at $\bf{x}$ and $1_{B_r}$ the associated
characteristic function. In principle, with infinitely many data points
and aribrarily fine sampling, one would consider the limit $r \rightarrow 0$.
In practice a suitable density estimator is needed.
Here we adopt a fixed-volume Epanechnikov kernel \cite{OL98,OL00},
with a compact support $B_r$ and a given finite radius $r$ (see below). 
As shown previously \cite{OL98,OL00}, smoothness of this
measure along the trajectory is a good candidate to quantify 
determinism. Numerically, the differentiability of the measure along
the trajectory was evaluated by means of the topological 
statistics developed by Pecora {\it et al.} \cite{PC95}. 
Actually we look at the continuity of the logarithmic
derivative of the measure (see \cite{OL00}).
 
We want to emphasize that in looking at this property of the measure
we avoid one of the most common problems found in this kind of algorithms,
namely, the lack of an acceptable scaling region. In fact, this is
the main problem encountered when dealing with quantities such
as the maximal Lyapunov exponents and/or  Kolmogorov entropy 
\cite{CF00}, particularly in the case of high dimensional systems
(see ref. \cite{KS97}).
Especially troublesome is the calculation of the information ($D_1$)
or correlation ($D_2$) dimensions,
in which the search for a reliable scaling region is the
source of most of the misuse of the algorithm. Contrarily,
our algorithm needs a realible estimation of
the natural measure on the attractor not in a range of scales
but at a fixed one. There is a lower value of resolution given
by the minimun average interpoint distance, which can be estimated
as $r_{min} \sim N^{-2/D_2}$, and an upper limit given by 
the attractor extent (which
we always normalize to the unit hyper-square) \cite{DG93}. Below the lower
limit there would be statistical fluctuations due to the lack
of neighboring points, while close to the upper limit we must confront
edge effects \cite{DG93,GM98}. The latter are particularly important for
high embedding dimensions because, as the dimension increases, 
more points stay near the attractor boundary. 
These bounds are by no means exact, for instance
the expression for the lower limit was derived by assuming  the 
data points to be uniformly distributed over
the attractor, which is not always true even in cases
as simple as nonlinear oscillators. In practice one should look in each 
particular case for the most appropriate scale to fix.
As far as the systems discussed here are concerned, we have found
that setting $r$ at 10\% of the attractors' linear extent is a good choice. 
 
\subsection{Statistical tests of continuity}
\label{Ss_Stc}

A naive test to quantify noise in signals is to
check how smooth they are. As long as more noise contaminates the signal,
more discontinuous it becomes. This is the case for example
of additive noise, {\it e.g.} noise added to the signal \cite{KS97,EO01}. 
However, this is by no means a general rule. For instance,
dynamical noise, that is, noise added in the equation of
motion, is not expected to affect the smoothness of the signal.
We  overcome this  drawback by using 
the distribution of points on the trajectory (or the natural
measure) as a way to evaluate the degree of
noise in the system (see \cite{OL00}). In
order to test the mathematical properties of the measure,
that is, continuity, differentiability, inverse
differentiability and injectivity, we borrow the statistical
approach developed by Pecora {\it et al.} \cite{PC95}.
Basically, the method is intended to evaluate,
in terms of probability or confidence levels, whether two data
sets are related by a mapping having the {\it continuity} property:
A function $f$ is said to be
continuous at a point ${\bf x}_0$ if $\forall \epsilon > 0, \exists
\delta > 0$ such that $\parallel {\bf x} - {\bf x}_0 \parallel <
\delta \Rightarrow \parallel f({\bf x})) - f({\bf x}_0) \parallel
< \epsilon$. The results are tested against the null--hypothesis,
specifically, the case in which no functional relation
between points along the trajectory and the measure exists.
Thus, as done in \cite{PC95}, we calculate
\begin{eqnarray}
\Theta_{C^0} (\epsilon) = \frac{1}{n_p} \sum_{j=1}^{n_p}
\Theta_{C^0} (\epsilon , j)
\label{e7}
\end{eqnarray}
and
\begin{eqnarray}
\Theta_{C^0} (\epsilon, j) = 1 - \frac{p_j}{P_{\rm max}}
\label{e8}
\end{eqnarray}
\noindent
where $p_j$ is the probability that all of the points in the
$\delta$-set, around a given point ${\bf x}_j$, of the reconstructed
trajectory,
fall in the
$\epsilon$-set around $\frac{d{\ln \mu({\bf x}_j)}}{dt}$.
The likelihood that this will happen must
be relative to the most likely event under the null hypothesis, $P_{\rm max}$
(see refs. \cite{PC95} and \cite{EO01}). When $\Theta_{C^0} (\epsilon, j)
\approx 1$ we can confidently reject the null hypothesis, and
assume that there exists a continuous function.
As in the work of Pecora {\it et al.} \cite{PC95} the $\epsilon$
scale is relative to the standard deviation of the density time series,
and thus, $\epsilon \in [0,1]$.
Plots of $\Theta_{C^0}(\epsilon)$ versus $\epsilon$ can be used to
quantify the degree of statistical continuity of a given function.
The typical outcome is a sigmoidal curve whose width and slope
are affected by the level and the type of noise contained in the
series \cite{OL98,OL00,EO01}.
In order to characterize the continuity statistics by means of
a single parameter we can also calculate,
\begin{equation}
\theta = \int_0^1 \Theta_{C^0} (\epsilon) {\rm d}\epsilon
\end{equation}
\noindent The limiting values of $\theta$, namely, 0 and 1, correspond
to a strongly discontinuous and a fully continuous function, respectively.
Hereafter we shall refer to $\theta$ as CS (Continuity Statistics).

\section{Results}
\label{S_Res}

\subsection{Generalized Mackey-Glass system}
\label{Ss_GMGs}

In order to investigate how our algorithm work on high
dimensional systems, we use a generalization of the 
Mackey-Glass (MG) equation \cite{HB98}, a delayed feedback 
system, 

\begin{eqnarray}
{\dot x(t)} & = & \frac{ax(t-\tau_0)}{1+x^{10}(t-\tau_0)}-y(t) \nonumber \\
{\dot y(t)} & = & -\omega^2 x(t) - \rho y(t)
\label{gMGe}
\end{eqnarray}
where $a = 3$, $\rho = 1.5$, $\omega = 1$ and $\tau_0 = 10$. 
As it is stated in ref. \cite{HB98} the Kaplan-Yorke 
dimension of this system is $D_{KY} \sim 13.5$, which, according
to the Kaplan-Yorke conjecture, $D_{KY} = D_1$, for a typical
attractor.
We have made a standard reconstruction analysis over 
time series of up to 16384 data points.
Although the optimal time delay $\tau$ should be
given by the first minimum of
the mutual information, $210$ in sampling units, we have
used a somewhat smaller value, typically $120$. Using
larger delays in high embedding dimensions would reduce drastically the
number of reconstructed points. The choice of this value, 
however, is supported by the fact that in high dimensions,
the optimal $\tau$ seems to be smaller than that given by
the mutual information criterion \cite{OK97}, at least in
the case of chaotic continuous-time systems, as our case.
Surrogate time series \cite{TE92,SS96}
with the same number of points have also been generated, using the
routines in the TISEAN \cite{HK99} package (``{\tt endtoend}" and 
``{\tt surrogates}").

We have also calculated the correlation integral of this system.
As it is well known, this is the most basic procedure to
discriminate between deterministic and stochastic behavior.
The numerical results for the correlation integral and its derivative
(that gives the correlation dimension) are shown in Figure 1. 
The results show that, for this  high dimensional system,
no region with constant slope is found. 
This clearly illustrates the well known difficulties inherent 
to the estimation of the correlation dimension in high dimensional systems.

In Figure 2 we report the values of CS for the generalized MG
system (reconstruction from the $x$-time series) when the
embedding dimension is varied in the range 2--30.
In order to increase the robustness of the results, we have averaged 
the CS for six differents time series.
The high values corresponding to the original series, together with 
the lower values of their surrogates,
indicate that our method is able to identify determinism in a
high dimensional system. We also note that
whereas the CS for the
original series remains almost constant for embedding dimensions
larger than the correlation dimension, that for the surrogate series
decreases steadily. Finally, it is here pertinent to observe
that the results of ref. \cite{OL00} are referred to
the standard MG equation (see
Eq. (\ref{stdMGe}) in the Appendix), whose correlation
dimension is estimated around seven \cite{KG92}, while
here we have choosed the generalization (\ref{gMGe})
just to check that the methodology works
with more severe test. However, despite the smaller
correlation dimension, in Fig. 8 of \cite{OL00}
the CS for the standard MG system is lower than
the CS for the generalized version reported here.
In the Appendix we discuss the solution to this
apparent paradox, through a digression on the role
of the sampling time on the observed levels of CS.
  
\subsection{Simple stochastic systems}
\label{Ss_Sss}

As already remarked above, the results just discussed have to be contrasted 
with those obtained for stochastic systems. Here we show results for
two time series derived from a linear (\ref{stoc1}) and a nonlinear 
(\ref{stoc2}) stochastic processes, namely, 

\begin{eqnarray}
{\dot x} & = & \theta x(t) + \eta (t)
\label{stoc1}
\end{eqnarray}
\begin{eqnarray}
{\dot y} & = & (\alpha - 0.5) \beta - y(t) + \sqrt{2 \beta y(t)} \eta(t)
\label{stoc2}
\end{eqnarray}
where $\eta(t)$ is a Gaussian noise with standard deviation 0.1, 
$\alpha = \beta = 1$, and $\theta = -0.9$.
Both processes are examples of purely stochastic systems, as no dynamics 
is left in the noise free limit, 
and exhibit $(1/f^{\alpha})$ power law spectra,
a finite correlation dimension ($\approx 2$) and a converging 
Kolmogorov entropy
\cite{OP89}. Note that this is a typical case in which the correlation 
dimension fails in identifying stochasticity. 

Attractor reconstruction was carried out on time series with up
to 8192 data points and a time delay of 100 (somewhat less than the
first minimum of the mutual information which in this case lies approximately
at 130). As it is clearly seen in Figure 3, there is no difference between the
CS for the orginal time series and its surrogate, both showing very
low values which indicate a low differentiability (a signature of 
stochasticity as discussed in \cite{OL00}). 

\subsection{Model of Electro-encephalogram signals}
\label{Ss_MEEGs}

A paradigmatic example of high dimensional systems
is found in physiology, namely, Electro-encephalogram (EEG) signals.
These are in fact the result of a sum over a large number of 
neuronal potentials. There are many studies that claim deterministic behavior
in EEG dynamics, most of them based on the calculation of the correlation
dimension. However, recent analyses have pointed out many technical problems
related to those studies (see \cite{JK99} and references therein)
which throw serious doubts on the above conclusion.
We have applied our method to  the analysis of a EEG-like signal
\cite{JK99} generated by the following set of nonlinear
coupled equations,
\begin{mathletters}
\begin{equation} 
{\dot x_1}=x_2
\end{equation} 
\begin{eqnarray} 
{\dot x_2}=\frac{x_5-25}{3}{\rm sin}\omega_1 t +3x_7{\rm sin}\omega_2 t  
\nonumber \\
+x_{11}{\rm sin}\omega_2 t -3|x_6|x_2-x_9x_1
\end{eqnarray} 
\end{mathletters}
\noindent with $\omega_1$=30, $\omega_2$=65 and $\omega_3$=80;
\begin{mathletters}
\begin{equation} 
{\dot x_3}=\sigma (x_4-x_3)
\end{equation} 
\begin{equation} 
{\dot x_4}=-x_3x_5+rx_3-x_4
\end{equation} 
\begin{equation} 
{\dot x_5}=x_3x_4-bx_5
\end{equation} 
\label{sLs}
\end{mathletters}
\noindent where $\sigma$=10, $r$=28 and $b$=8/3 (Lorenz system).
\begin{mathletters}
\begin{equation} 
{\dot x_6}=x_7
\end{equation} 
\begin{equation} 
{\dot x_7}=-kx_7-x_6^3+B{\rm cos}t
\end{equation} 
\end{mathletters}
\noindent where $k$=0.1 and $B$=12 (Ueda equations).
\begin{mathletters}
\begin{equation} 
{\dot x_8}=x_9
\end{equation} 
\begin{equation} 
{\dot x_9}=-\delta x_9+\frac{1}{2}x_8(1-x_8^2)+f{\rm cos}\omega t
\end{equation} 
\end{mathletters}
\noindent where $\delta$=0.15, $F$=0.15 and $\omega$=0.8 (two--well
potential Duffing--Holmes attractor).
\begin{mathletters}
\begin{equation} 
{\dot x_{10}}=-(x_{11}+x_{12})
\end{equation} 
\begin{equation} 
{\dot x_{11}}=x_{10}+\alpha_{11}
\end{equation} 
\begin{equation} 
{\dot x_{12}}=\alpha+x_{12}(x_{10}-\mu)
\end{equation} 
\end{mathletters}
\noindent where $\alpha$=0.15 and $\mu$=10 (R\"ossler attractor).
This intricate system has a correlation dimension around 9 
and has failed to pass the method originally proposed by Salvino and 
Cawley \cite{SC94}.

We have integrated the whole system of \cite{JK99} using a 
fourth-order Runge-Kutta algorithm with a fixed step of 0.001.
The main two coordinates, $x_1$ and $x_2$, behave in a markedly different way.
Coordinate $x_2$ shows a rather standard behavior with
the first zero of the autocorrelation function lying at 
$\simeq 27$ (in units of sampling time). 
Instead, the $x_1$ time series
exhibits a power spectrum  that seems
to be of the $1/f^2$ type and a first zero of the autocorrelation
that strongly depends on the series length $N$. We found
$1.3 \times 10^4$ for $N=10^5$ and
$9.9 \times 10^4$ for $N=10^6$ (both in units
of a larger time step, namely 0.01, that was used
to analyze longer time series).
These results may reveal a non-stationary behavior of $x_1$. In fact this 
coordinate shows a saw-tooth shape with a very long wavelength and, 
as a consequence, it may appear to increase linearly over rather long 
time intervals. 
As the analysis of \cite{JK99} was done on the $x_1$ coordinate it may
actually be the reason why those authors failed in detecting
determinism in this system. 
Note that the lack of stationarity (a property related to ergodicity) would 
imply the failure of our method. A proper estimation of the
(reconstructed) measure, requires that space averages 
correspond to time averages, which in fact would be impossible
if the time series is not stationary. This is the reason why we
cannot estimate confidently the measure along the trajectory
using $x_1$ in the above system. However, as ${\dot x_1} = x_2$,
this allow us a better estimate from the 
$x_2$ coordinate (since upon differentiating \cite{SM90} the saw--tooth
shape of the $x_1$-time series coordinate results in very small offsets).

Figure 4 shows the CS for both the $x_2$ time series and its surrogate.
We have averaged over 8 differents realizations and their
corresponding surrogates.
The results correspond to time series having 4096 data
points and a time delay of 27 (in this case the first zero of
the autocorrelation function and the first minimum of the mutual information
almost coincide). 
The difference between the CS for the original series and that for its
surrogate is noticeable, even on such short time series. 
This, along with the high value of CS obtained
for the original series \cite{OL00}, reveals an essentially deterministic 
origin of the simulated EEG signal. 

\section{Discussion}
\label{S_D}

The examples discussed in the previous section indicate
that the analysis of the CS with respect to the embedding
dimension reveals useful informations in order to discriminate 
stochasticity from determinism. Despite the high dimensionality
of the deterministic systems we notice that the distinction
is feasible almost from the start, that is, from dimensions lower than the
attractor dimension. This is an interesting (though somehow
``fortunate'') outcome since, for such low embedding dimensions, 
transverse self-intersections 
of the orbit must occur (or equivalently the existence of false nearest
neighbors). 
However, as the embedding dimension
increases, and especially beyond the correlation dimension of the
system, the CS of the original signal seems to stabilize around a 
constant value.
Contrarily, the CS for the surrogated series steadily decreases 
for embedding dimensions larger than the correlation dimension
(compare Figures 2 and 4). Finally, the CS for the purely
stochastic processes (Figure 3) attains rather low values
and is monotonically decreasing.

The small error bars in the CS, both in Figure 2 and 4, indicate that
the smoothness of the measure in deterministic systems shows almost no
dependence on initial conditions. The absolute
value of the CS for deterministic signals may slightly depend on several
factors, namely, number of data points, sampling rate, ball
size in the measure estimation process, etc. But, for 
most deterministic time series belonging to different trajectories,
the computed CS seems to be a robust measure of the 
property we want to identify. On the other hand, the CS calculated for the
reconstructed measure in the case of the surrogate time series
shows a large variability. In noting this we are not only referring to the 
large error bars of Figures 2 and 4 characteristic of most noisy magnitudes, 
but rather to the largely different CS found for the systems 
of those two Figures. In particular we note  that, while the CS for 
the original series 
are similar and higher than 0.9, those for the surrogated series 
are, higher than 0.6 in the case of the generalized Mackey-Glass (Fig. 2) 
and  smaller than 0.2 for the EEG signal (Fig. 4). 
Surrogate time series generated by the IAAFT method \cite{SS96}, as  
those used here, are in fact, realizations of a {\it Gaussian linear
stochastic process} (possibly passed through a nonlinear
invertible measurement function). By definition, they share the same 
correlation structure of the original time series and
the same probability density function (histogram).
In passing, notice that in refs. \cite{OL98,OL00} the results
correspond to the simpler AAFT method.
Although correlation structure is in some
way related to continuity, both are very different 
properties of the signal. A strict application of the
surrogate methodology in our case would requiere to generate
surrogates time series with the same continuity characteristics of
the original signal. Moreover, and going back to the issue 
of the absolute value of the CS, we should keep in mind 
that the statistics of Pecora, Carroll and Heagy \cite{PC95} are in 
turn a measure of a certain null-hypotesis. Strictly speaking
we are dealing with the superposition of two null-hypotesis and
this drawback doesn't allow a clear interpretation
of the CS values coming from surrogate data.
It is likely that a different surrogation method is required
for a safe application of our method to the analysis of surrogate series
derived from nonlinear systems.  We are actually investigating
this issue in the light of constrained randomization \cite{Sc98}.

\section{Concluding Remarks}
\label{S_CR}

Summarizing, we have shown numerically that it is possible to
discriminate determinism from stochasticity in high dimensional
systems. A local property of the system, as the differentiablity
of the measure along the trajectory, prevents us from regarding as "random"
those systems that actually are deterministic (a task in which other
methods have failed). The fact that for sufficiently fine sampling
and large embedding dimensions the parameter CS for deterministic
systems is always above 0.9, opens the possibility of applying our
methods to the analysis of more complicated series such as the 
experimental ones without resorting to  comparison with the CS for surrogate
series.  A lot of work still remains to 
establish the method as a truly practical tool. For instance, its limits
of applicability have to be identified,
particularly regarding the relation between the minimum 
number of data points in the series and the actual dimension of the
system. Real applications would require to extract information
about the existence of many more (hundred, thousand, etc.) 
active degrees of freedom of the underlying system. Fortunately, 
as the results here discussed seem to indicate, 
this does not necessarily require
large embedding dimensions. 

\acknowledgements
We aknowledge the freely available package TISEAN which we have used.
This work was supported by grants of the spanish CICYT (grant no. PB96--0085),
the European TMR Network-Fractals c.n. FMRXCT980183,  
the Universidad Nacional de Quilmes (Argentina) and the Universidad de
Alicante (Spain). G. J. Ortega is a member of CONICET Argentina.

\section{Appendix: How results depend on sampling}
\label{S_App}

One of the very preliminar steps in time series analysis
in to ensure that one is dealing with a good sampling time.
The precise meaning of this statement depends somehow
on the specific problem at hand. Nonetheless, what is
normally done is to check that the sampling time is
much smaller than the smallest time scale, $t_{\rm min}$,
present in the system. This could be a (quasi) period,
an autocorrelation time, an inverse of the maximum
Lyapunov exponent and so on. However, there are practical
limitations to this rule. Experimentally, the resolution
time is not always freely adjustable. Numerically, one
should cope with the limited performances of the
machine. In both cases, one has to consider also
the problem of storing large amount of data.
This is fundamental when one realizes that
the system possesses intrinsic long time scales,
which have to be covered in order to analyze
{\it stationary} time series.
The compromise between series length and
stationarity is often achieved by making
a re-sampling of the series itself.
However, this decimation procedure is not always innocuous.
This Appendix is devoted to study how our basic
index, the CS, is affected when one takes
different sampling time.
We take the following three examples. First,
time series
built from the $x$-coordinate of the Lorenz
system ($x_3$ in Eqs. (\ref{sLs})).
Second, 
the Mackey-Glass delayed differential
equation:
\begin{equation}
\dot{x}=\frac{a x(t-\tau_0)}{1+x^{10}(t-\tau_0)}-b x(t) \; ,
\label{stdMGe}
\end{equation}
choosing $a=0.2$, $b=0.1$ and $\tau_0=100$. Third, the purely
(nonlinear) stochastic system of
Eq. (\ref{stoc2}). 
In Fig. 5 we report the corresponding
CS's versus the embedding dimension, for various
values of the decimation time. The resulting series
lengths are 10000 for the Lorenz and MG systems
and 20000 for the stochastic process.
In every case, the delay
time of the reconstruction was choosen as the
minimum of the corresponding mutual information.
As far as the Lorenz and MG systems
are concerned, one sees what it is somehow expected
for a discretized continuous function, that is,
the coarser is the sampling the lower is the continuity level.
Hence, when applying this kind of methods, one should
keep in mind that to some extent the CS may be lowered
due to a not-so-good sampling. In particular from
Figs. 2 and 5 one sees that is in fact the origin
of the discrepance, anticipated in Subsection \ref{Ss_GMGs},
between the MG data shown
here and those of ref. \cite{OL00}.
Nevertheless, an important feature must be pointed out, namely that
this ``runoff'' is not arbitrary. The CS for the determinstic
systems doesn't fall down to the values corresponding to
the stochastic system which in turn appear to be essentially
independent of the sampling time. It is interesting to
speculate to what extent this last property can be exploited
to build up a technique based on re-sampling, where
the stochastic
signals are identified just as those which are ``squeezed down''
to rather low values of CS, independently of their resolution.


\begin{figure}[tbh]
\begin{center}
\
\psfig{file=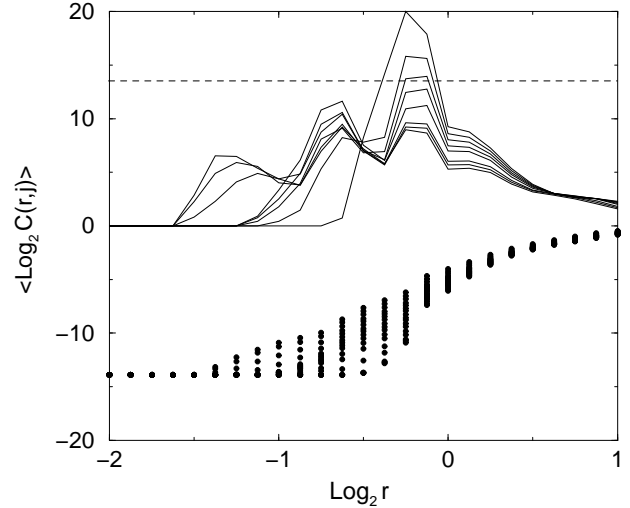,height=3.in}
\caption{Log-Log plot of the correlation integral, with
embedding dimensions in the range 10 to 40, for the $y$--coordinate of 
system (\ref{gMGe}). 16384 data points has been used.
Upper lines show the derivative of the correlation integral for
embedding dimensions of 20, 22, 24, 28, 30, 32, 34 and 36.
Horizontal dashed line shows the actual information dimension of the
system.}
\end{center}
\end{figure}

\begin{figure}[tbh]
\begin{center}
\
\psfig{file=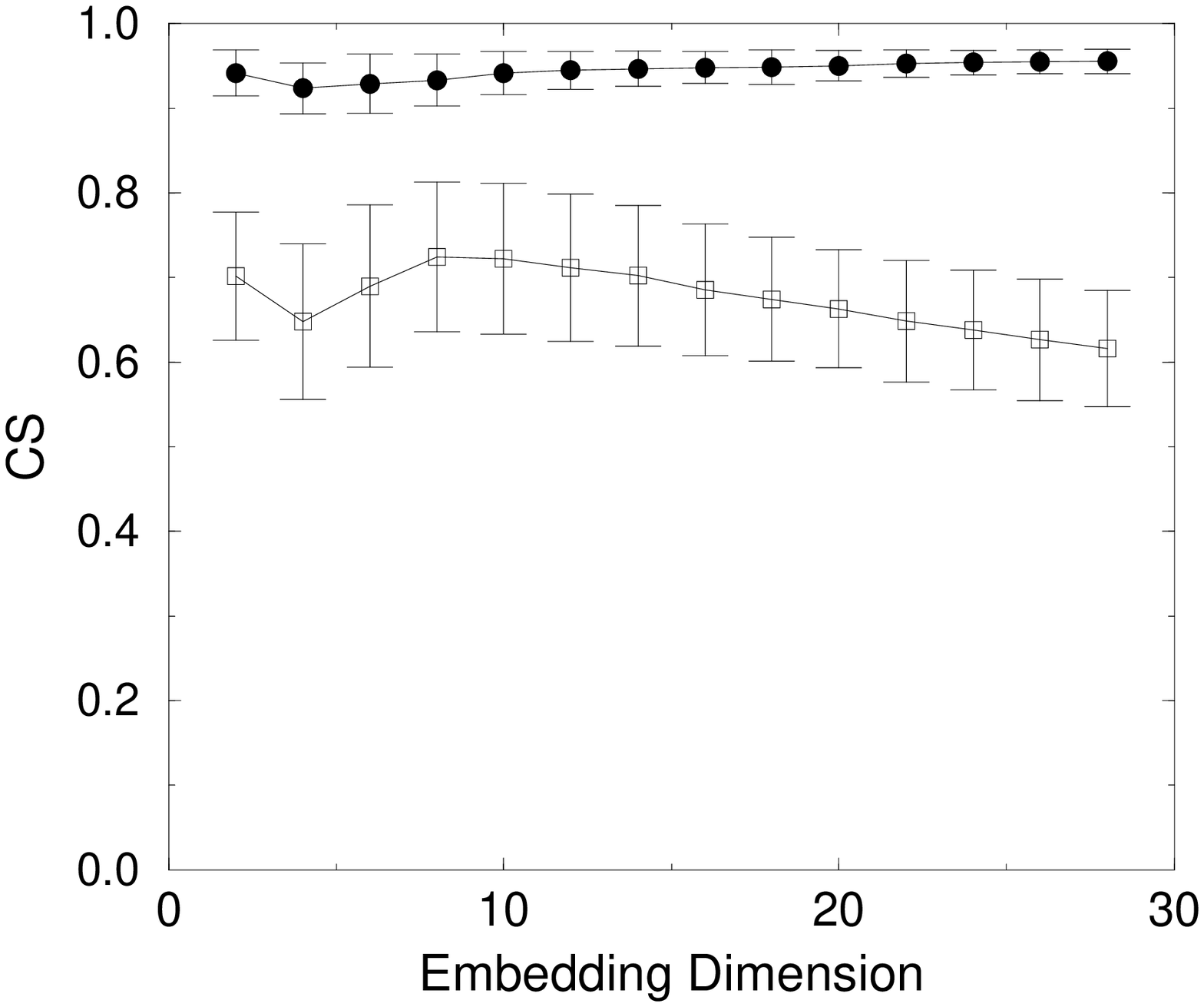,height=3.in}
\caption{CS for the system (\ref{gMGe}) (filled circles), 
averaged over six differents time series 
of 16384 data points each. Error bars (standard deviation) are also shown.
The $y$--coordinate has been used.
The empty symbols correspond to their respective surrogates.}
\end{center}
\end{figure}

\begin{figure}[tbh]
\begin{center}
\
\psfig{file=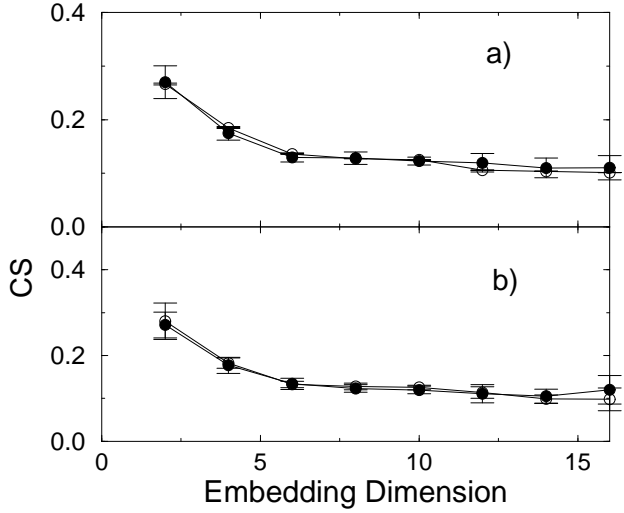,height=3.in}
\caption{CS for random signals. a) Solid circles correspond
to Eq. (\ref{stoc1}) and empty circles are the corresponding
surrogates (differences are not visible as the points almost coincide). 
Eight differents realizations of 8192 each have
been used. b) Idem as a) but for Eq. (\ref{stoc2}).}
\end{center}
\end{figure}

\begin{figure}[tbh]
\begin{center}
\
\psfig{file=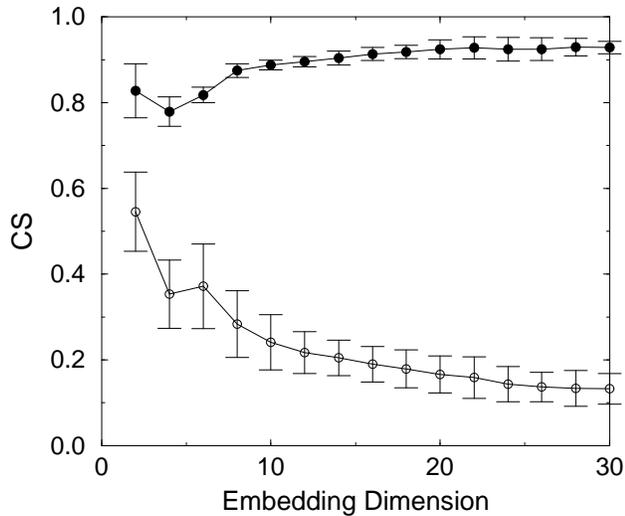,height=3.in}
\caption{CS for the EEG-like signal. The results  correspond
to the $x_2$ coordinate of the system. 4096 data points 
were used in each of the eight time series averaged.
The  standard deviation (error bars) is also shown. 
Empty symbols correspond to their respective surrogates.}
\end{center}
\end{figure}

\begin{figure}[tbh]
\begin{center}
\
\psfig{file=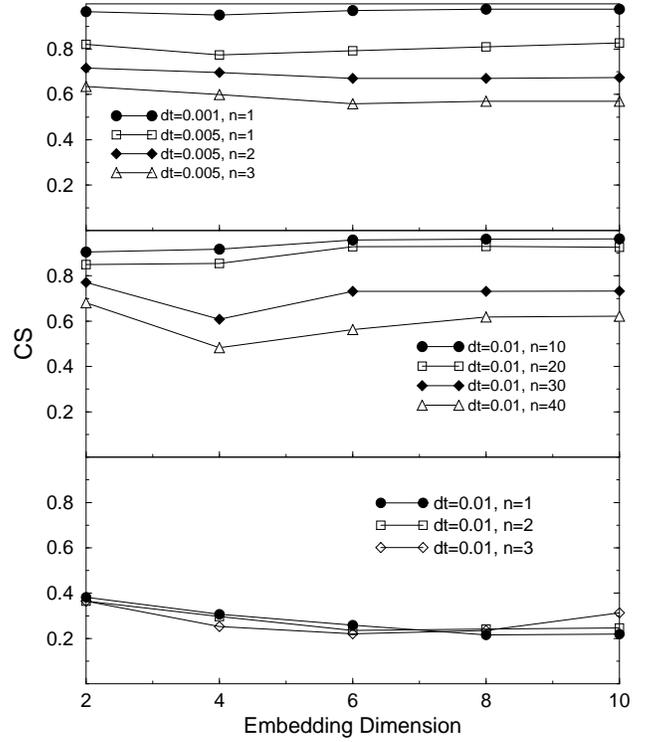,height=4.in}
\caption{Dependence of the CS on the sampling time for
different continuous-time systems. Upper panel: Lorenz
system; Central panel: Standard Mackey-Glass equation
(Eq. (\ref{stdMGe})); Lower panel: Random system
of Eq. (\ref{stoc2}). The symbol {\tt dt} indicates 
the time step used in the numerical
integration scheme (fourth-order Runge-Kutta for
the first two and Euler's for the random system)
while $n$ corresponds to the re-sampling step.}
\end{center}
\end{figure}


\begin{references}

\bibitem{SM90}
G. Sugihara and R. May, Nature (London) {\bf 344}, 734 (1990).
\bibitem{KG92}
D.T. Kaplan and L. Glass, Phys. Rev. Lett. {\bf 68}, 427 (1992);
Physica D {\bf 64}, 431 (1993).
\bibitem{WB93}
R. Wayland, D. Bromley, D. Pickett and A. Passamante,
Phys. Rev. Lett. {\bf 70}, 580 (1993).
\bibitem{SC94}
L.W. Salvino and R. Cawley, Phys. Rev. Lett. {\bf 73}, 1091 (1994).
\bibitem{BK99}
J. Bhattacharya and P. P. Kanjil, Physica D {\bf 132}, 100 (1999).
\bibitem{KS97}
H. Kantz and T. Schreiber, {\it Nonlinear Time Series Analysis}
(Cambridge University Press, Cambridge, 1997).
\bibitem{AB93}
H. Abarbanel, R. Brown, J. Sidorowich and L. Tsimring,
Rev. Mod. Phys. {\bf 65}, 1331 (1993).
\bibitem{CF00}
M. Cencini, M. Falcioni, E. Olbrich, H. Kantz and A. Vulpiani,
Phys. Rev. E {\bf 62}, 427 (2000).
\bibitem{JK99}
J. Jeong, M.S. Kim and S.Y. Kim,
Phys. Rev. E {\bf 60}, 831 (1999).
\bibitem{CE91}
M. Casdagli, S. Eubank, D. Farmer and J. Gibson,
Physica D {\bf 51}, 52 (1991).
\bibitem{OL98}
G. J. Ortega and E. Louis, Phys. Rev. Lett. {\bf 81}, 4345 (1998).
\bibitem{OL00}
G. J. Ortega and E. Louis, Phys. Rev. E. {\bf 62}, 3419 (2000).
\bibitem{PC95}
L. M. Pecora, T. L. Carroll and J. F. Heagy, Phys. Rev. E. {\bf 52}, 3420 (1995).
\bibitem{DG93}
M. Ding {\it et al.} Physica D {\bf 69}, 404 (1993).
\bibitem{GM98}
A. Galka, T. Maab and G. Pfister,
Physica D {\bf 121}, 237 (1998).
\bibitem{EO01}
C. Degli Esposti Boschi, G. Ortega and E. Louis, submitted to Physica D.
\bibitem{HB98}
R. Hegger, M. J. B\"{u}nner, H. Kantz and A. Giaquinta, 
Phys. Rev. Lett. {\bf 81},
558 (1998).
\bibitem{OK97}
E. Olbrich and H. Kantz, Phys. Lett. A {\bf 232(1-2)}, 63 (1997).
\bibitem{TE92} J. Theiler, S. Eubank, A. Longtin, B.
Galdrikian and J. D. Farmer, Physica D {\bf 58}, 77 (1992).
\bibitem{SS96} T. Schreiber and A. Schmitz, Phys. Rev. Lett.
{\bf 77}, 635 (1996).
\bibitem{HK99}
R. Hegger, H. Kantz and T. Schreiber,
Chaos {\bf 9}, 413 (1999).
\bibitem{OP89}
A. Osborne and A. Provenzale, Physica D {\bf 35}, 357 (1989).
\bibitem{Sc98}
T. Schreiber, Phys. Rev. Lett {\bf 80}, 2105 (1998).

\end{references}
\end{document}